\documentstyle[12pt,epsf]{article}
\begin{document}
\bibliographystyle{unsrt}
\baselineskip=12pt 

\title{Density of Topological Defects After a Quench}

\author{Pablo Laguna${}^{(1)}$ and Wojciech Hubert Zurek${}^{(2)}$\\
\\
${}^{(1)}$ Department of Astronomy \& Astrophysics and \\
Center for Gravitational Physics \& Geometry\\
Penn State University, University Park, PA 16802, USA\\
email: pablo@astro.psu.edu\\
\\
${}^{(2)}$ Theoretical Astrophysics, MS-B288\\
Los Alamos Nat. Lab., Los Alamos, NM 87545, USA\\
email: whz@lanl.gov}

\date{\today}
\maketitle

\begin{abstract}
We present results of numerical studies of the Landau-Ginzburg
dynamics of the order parameter in one-dimensional models 
inspired by the condensed matter analogues of cosmological phase transitions.  
The main goal of our work is to show that, as proposed by
one of us \cite{Zurek85b}, the density of the frozen-out topological defects is set by 
the competition  between the quench rate --- the rate at which 
the phase transition is taking place --- and the relaxation rate of 
the  order parameter.  In other words, the characteristic domain size,
which determines the typical separation of topological defects in 
the new broken symmetry phase, is of the order of the correlation length 
at the instant at which the relaxation timescale of 
the order parameter equals the time remaining to the phase transition.  
In estimating the size of topological domains,
this scenario shares with the original Kibble mechanism
the idea that topological defects will form along the 
boundaries of independently selected regions of the new broken
symmetry vacuum. However, it derives the size
of such domains from non-equilibrium aspects of the
transition  (quench rate), as opposed to Kibble's
original proposal in which their size was  
estimated from the Ginzburg temperature above 
which thermally activated symmetry restoration can occur.
\end{abstract}

\section{Introduction}

The expansion of the Universe inevitably leads to a drop of the temperature of
the primordial fireball.  This cooling provides natural conditions that
are believed to precipitate a series of phase transitions 
in which ``false'' symmetric, high temperature phases are transformed
into low temperature broken symmetry ``true'' vacuum.
As the Universe  undergoes such phase
transitions, the selection of the low-temperature, broken symmetry phase can only occur 
locally, within the causally correlated regions.   
Kibble \cite{Kibble76,Kibble80} first 
noted that this symmetry breaking process may leave
relics of the high energy phase which will be
trapped by the topologically stable configurations of the broken symmetry  phase.  
In principle, such topological defects could be massive enough to leave an 
observational imprint 
in the cosmic microwave background and perhaps also produce the seeds needed 
for matter structure formation \cite{Kibble76,Zeldovich75}.

There are three principal types of topological defects \cite{Vilenkin94}.
They differ by their dimensionality.  Monopoles are point-like objects 
that would be disastrous from the cosmological point of view.
They have a tendency to dominate the matter content of the Universe.  
Inflation was invented in part to dilute them and thus to 
prevent observational inconsistencies.
Membranes or domain walls are two-dimensional, and they
almost certainly could also lead to the same disaster as monopoles.   
Specifically, they would induce unacceptably large distortions of the
cosmic microwave background.  
By contrast, one-dimensional topological defects, 
cosmic strings, could provide the source of
density perturbations needed to seed structure formation \cite{Kibble80}.

Symmetry breaking phase transitions which occur in condensed 
matter physics are described by theories which 
are formally analogous to those involved in the cosmological context
(for a review see \cite{Zurek96}).
Condensed matter phase transitions have, however,
one crucial advantage:  they can be 
studied in the laboratory.  With this in mind, more than a decade ago
one of us suggested \cite{Zurek85b,Zurek85a} 
that the cosmological  mechanism for defect formation can be
studied experimentally in the condensed matter context.  
A number of experiments have been carried out following that proposal.
For instance, liquid crystal experiments demonstrated that copious production of 
topological defects does indeed happen in symmetry breaking phase
transitions \cite{Chuang91,Bowic94}. 
More recently, superfluid He$^4$ and He$^3$ experiments have also 
tested the topological defect formation picture \cite{Hendry94,Ruutu96,Bauerle96}.
Liquid crystal phase transitions, however, are of the first order type.
In first order phase transitions, 
the process which is responsible for the appearance of the new phase is 
nucleation. Namely, small regions of the medium undergo thermal 
activation which takes them over
the potential barrier  separating ``false'' and ``true'' vacua.  
As a result, bubbles of a certain
(critical) size appear and form seeds of the  new phase.  
Eventually, through growth and
coalescence of these bubbles, the new phase replaces the old one.
A much more interesting and less understood dynamics takes 
place in second order  (Landau-Ginzburg like) phase transitions.
For this reason, He$^4$ and He$^3$ systems, for which
the transition to a superfluid phase proceeds without nucleation, 
have recently attracted significant interest as examples of cosmological
phase transition analogues. 

The defect formation scenario is based on two assumptions.  
The first assumption is of the qualitative nature:  
It asserts that regions of the broken symmetry phase,
which are  causally disconnected, must select the new low temperature phase
independently. As a result, when 
a symmetry-breaking phase transition  with a non-trivial
vacuum manifold occurs simultaneously in a sufficiently 
large volume, topological defects will 
be formed.  The second assumption is of the quantitative nature:  It
involves  specifying the process 
responsible for the causal propagation of ``signals''
that allow the choice of  the new vacuum 
to occur in a coordinated (rather than independent)
fashion.  This is the assumption that yields a 
prediction of the density  of topological defects. 
Both assumptions are of course necessary \cite{Kibble76}, 
but while the first one is straightforward,
the second one requires much more specific physical input. 

The original discussion of the scenario for defects formation was based on an
idea similar to that of nucleation 
in first order phase transitions \cite{Kibble76}.  It
was thought that the initial density of the 
topological defects is set at the so-called
Ginzburg temperature $T_G$, namely the temperature below the critical temperature
$T_C$ at which thermally activated  transitions 
between the correlation-sized volumes of
the new broken symmetry phase cease to occur. Below $T_G$, the
(free) energy barrier, separating ``false'' and ``true'' vacua,
becomes prohibitively large for correlation-length sized thermal
fluctuations.  Under this approach then, the
density of defects is set by the correlation
length at $T_G$.  

One of the key predictions of the original papers \cite{Zurek85b,Zurek85a}
on ``cosmological'' phase
transitions in superfluids was that  
this thermally activated process \underline{does not}
decide the density of defects.
Instead it was proposed  that the characteristic correlations 
length is set by the dynamics of the order
parameter in the vicinity of the  critical temperature $T_C$.
This ``non-equilibrium'' scenario \cite{Zurek85b,Zurek85a} represents 
a significant change of the point of view, and yields a 
prediction for the density of defects which is rather different 
from the ``equilibrium'' estimate based on the Ginzburg temperature.  

\section{Symmetry Breaking} 

In the vicinity of the critical temperature $T_C$ of 
a second-order phase transition, the contribution 
from the potential $V$ to the free energy of the system is given by
\begin{equation}
V = \frac{\lambda}{8}(|\Psi|^4 - 2\sigma^2\epsilon|\Psi|^2 + \sigma^4\epsilon^2),
\label{eq:1}
\end{equation}
with $\lambda$ and $\sigma$ positive parameters and 
$\epsilon  = (T_C-T) /T_C$ the dimensionless
temperature parameter. The last term in (\ref{eq:1}) has been added, so 
$V=0$ at the minimum of the potential for $T \le T_C$.
Here $\Psi$ represents the \underline{order parameter}, 
an abstract measure of the degree to which the 
symmetry in question has been broken.

Above the critical temperature, the potential (\ref{eq:1}) has a single
minimum at $|\Psi| = 0$. 
As the temperature drops below $T_C$,
the shape of the potential changes; $\epsilon$ becomes non-negative.  
Instead of a single minimum at $|\Psi| = 0$, the potential has 
a minimum value $V=0$ at $|\Psi| = \sigma\sqrt{\epsilon}$. 
At the same time, the potential $V$
develops a barrier  centered at $|\Psi| = 0$ with magnitude
\begin{equation}
\Delta V = V(0)-V(\sigma\sqrt{\epsilon}) = \frac{1}{8}\lambda\sigma^4\epsilon^2 \, , 
\label{eq:dv}
\end{equation}
``separating" the minimum values of the potential.

Depending on the symmetry that is broken, the manifold of equivalent
vacuum states for the order parameter will have different topologies,
and thus describe a different set of topological defects \cite{Vilenkin85}.
For instance, in the case of domain walls, the order parameter is a real scalar field
with discrete $\Psi \rightarrow -\Psi$ symmetry.
The vacuum manifold in this case consists of 
the two values $\Psi = \pm \sigma\sqrt{\epsilon}$.
Simple models of cosmic strings are, on the other hand, represented by a 
a complex scalar field with $U(1)$ symmetry of phase transformations,
$\Psi \rightarrow \hbox{e}^{i\alpha}\Psi$.
The degenerate vacuum states are given by 
$\Psi = \sigma\sqrt{\epsilon}\,\hbox{e}^{i\theta}$,
with $\theta$ an arbitrary phase. Thus, the vacuum manifold for 
cosmic strings is the circle $|\Psi| = \sigma\sqrt{\epsilon}$
in the complex $\Psi$-plane.

As mentioned before, since all the points in the vacuum manifold 
are completely equivalent, the choice of the expectation value of
the order parameter is determined from 
the initial thermal (random) fluctuations through
its dynamics. A measure of the extent, in space and time, to which perturbations
of the order parameter return to their equilibrium value
is given by the correlation length
$\xi$ and dynamical relaxation time $\tau$.
Specifically, the correlation function of perturbations away from equilibrium
behaves as
\begin{equation}
\langle \delta \Psi(x,t), \delta \Psi(x + \Delta,t)
\rangle  \sim \exp(-|\Delta|/\xi)
\end{equation}
and 
\begin{equation}
\langle \delta \Psi(x,t), \delta \Psi(x,t + \Delta)
\rangle  \sim \exp(-|\Delta|/\tau) \, .
\end{equation}
Furthermore, in the vicinity of the critical temperature $T_C$, 
the equilibrium values of $\xi$ and $\tau$ are given by
\begin{equation}
\xi  =  \xi_0\, | \epsilon |^{-\nu}
\label{eq:xi}
\end{equation}
and 
\begin{equation}
\tau  = \tau_0\,| \epsilon |^{-\mu} \, ,
\label{eq:tau}
\end{equation}
respectively, with $\mu$ and $\nu > 0$.
That is, at the critical temperature ($\epsilon = 0$),
the order parameter has a universal behavior characterized by the 
simultaneous divergence of both $\xi$ and $\tau$. 

\section{Freeze-out of Topological Defects}

Before presenting the results of the density of topological defects
from our numerical study, let us review the original estimate
of the density of defects put forward in \cite{Zurek85b}.
This estimate was based on a linear quench,
\begin{equation}
\epsilon = \frac{(T_C-T)}{T_C} = \frac{t}{\tau_Q} \, ,
\label{eq:quench}
\end{equation}
with $t$ the time before ($t<0$) and after ($t>0$) the transition.
(The idea behind the argument does not rely on this particular time dependence 
of the temperature parameter $\epsilon$, but it seems to be a likely
approximation when narrow range of temperatures near critical $T_C$ is
involved, as it will be the case here.) In (\ref{eq:quench}),
$\tau_Q$ represents the quench timescale. The other relevant timescale 
--- which we shall call symmetry breaking timescale --- is defined
by $|\epsilon/\dot\epsilon| = |t|$.
It is then possible to identify
during a quench three regimes  by comparing the magnitude
of the symmetry breaking timescale $|t|$ with 
the dynamical relaxation time $\tau$.

\begin{figure}[h]
\leavevmode
\epsfxsize=4.0truein\epsfbox{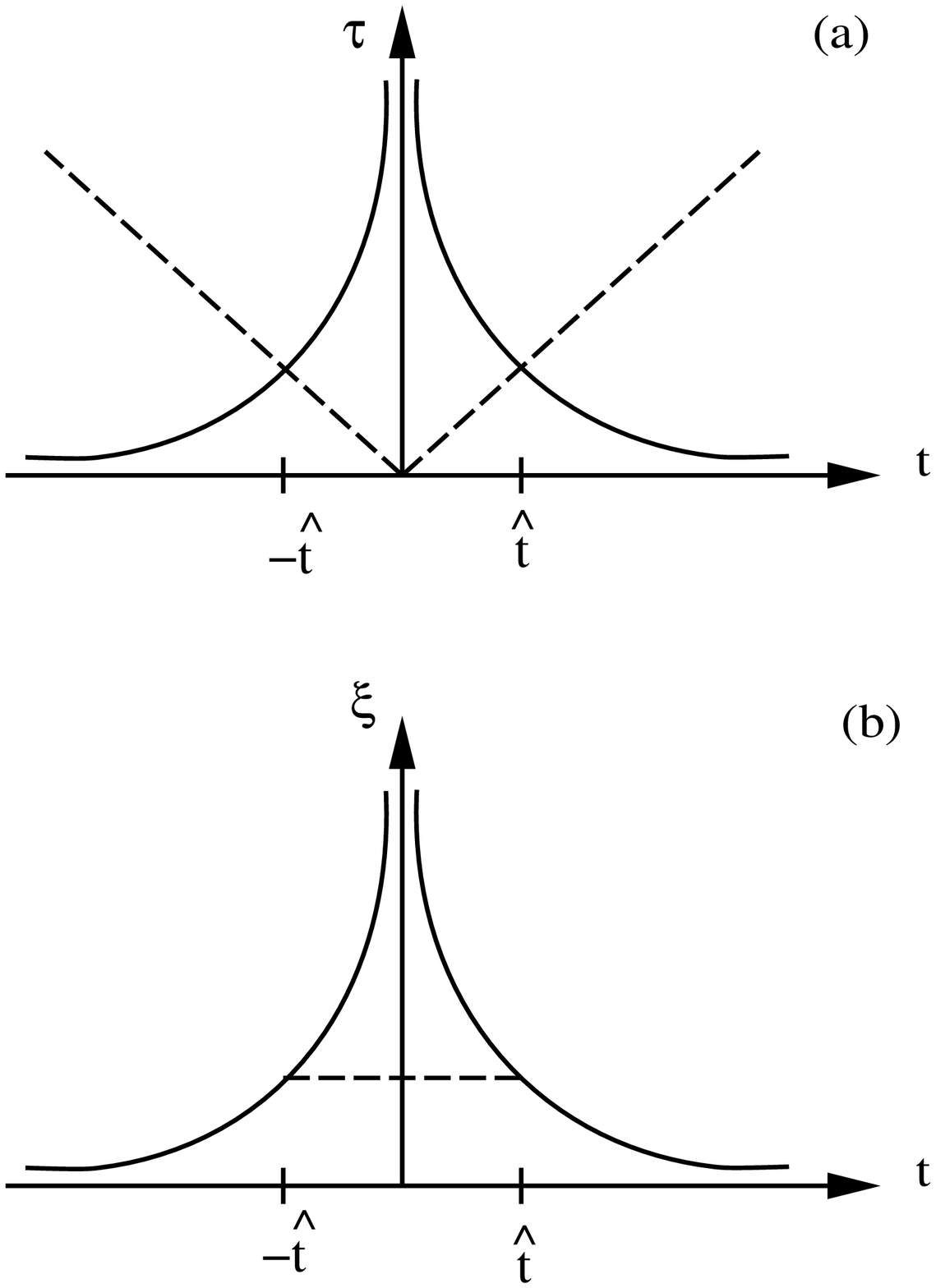}
\caption[figure1]{\label{figure1}
Characteristic behavior of the relaxation time ($a$) and the correlation 
length ($b$) during a quench. Dashed lines in ($a$) represent the symmetry breaking
timescale $|\epsilon/\dot\epsilon| = |t|$. Dashed line in ($b$) 
represents the freeze-out correlation length during the time interval
[$-\hat t, \hat t$].}
\end{figure}

For sufficiently large values of the temperature above $T_C$
(see Fig.~\ref{figure1}), 
the symmetry breaking timescale $|t|$
is larger than the dynamical relaxation time $\tau$.
The system is in a stage where the order parameter is able to 
adjust to the changes implied by the quench in an essentially adiabatic fashion.
In particular, the non-equilibrium correlation length closely matches
the equilibrium behavior given by Eq.~(\ref{eq:xi}).

Because of the critical slowing down implied by Eq.~(\ref{eq:tau}),
at some time $-\hat t$ before the transition,
the systems reaches the point at
which the symmetry breaking timescale $|t|$
and the dynamical relaxation time $\tau$ are comparable; namely
\begin{equation}
\tau(-\hat t)  = \tau_0 \,\Bigl( \frac{\tau_Q}{\hat t}\Bigr)^{\mu}  = \hat t\, .
\label{eq:hatt}
\end{equation}
The value of the temperature parameter at this point is
\begin{equation}
\hat\epsilon \equiv \epsilon(-\hat t) = 
\Bigl( \frac{\tau_o}{\tau_Q}\Bigr)^{\frac{1}{1+\mu}} \, .
\label{eq:ehat}
\end{equation}
The corresponding value of the correlation length can be directly found
by substituting Eq.~(\ref{eq:ehat}) into Eq.~(\ref{eq:xi}), 
\begin{equation}
\xi(-\hat t) = \xi_o \,
\Bigl(\frac{\tau_Q}{\tau_o}\Bigr)^{\frac{\nu}{1+\mu}} \, .  
\label{eq:xihat}
\end{equation}
Within the $[-\hat t,\hat t]$ interval,
the system is incapable of keeping up with the changes induced by the quench. 
In particular, the non-equilibrium correlation length cannot longer
grow at the equilibrium rate implied by Eq.~(\ref{eq:xi}).
Crudely speaking, the non-equilibrium correlation length
``freezes out" at the value $\xi(-\hat t)$ in Eq.~(\ref{eq:xihat})
until a time $\hat t$, after the transition, when once again
the dynamical relaxation time $\tau$ becomes smaller that 
the symmetry breaking timescale $|t|$. That is, approximately,
\begin{equation}
\xi(-\hat t) \approx \xi(\hat t) \equiv \hat\xi.
\end{equation}
The actual behavior of $\xi(t)$ is probably more complicated
(i.e. correlation will continue to grow even within the 
$[-\hat t,\hat t]$ interval), but $\hat\xi$ does capture the size
of the correlated region of vacuum at the time when its growth falls
out of step with the equilibrium scaling.
The bottom line is that during the interval [$-\hat t, \hat t$], the
order parameter, and therefore the correlation length,
is to sluggish to keep up with Eq.~(\ref{eq:xi}).

The proposal in \cite{Zurek85b,Zurek85a} states 
that the frozen out correlation length $\hat\xi$
yields the density of topological defects.
That is, $\hat\xi$ characterizes the size of independently selected
domains of the new vacuum, so it determines the separation 
(and therefore density) of topological defects.
As mentioned before, the difference with the original  
Kibble mechanism \cite{Kibble80} is the role that non-equilibrium
dynamics play in determining the value of $\hat\xi$.
In that picture, $\hat\xi$ was obtained
entirely from the temperature $T_G$ (Ginzburg temperature) at which 
fluctuations between the degenerate minima of the potential ceased to be possible;
that is, the temperature at which 
\begin{equation}
k_BT \approx \hat\xi^3 \Delta V
\end{equation}
with $\Delta V$ the potential ``barrier" 
in Eq.~(\ref{eq:dv}) separating degenerate minima.

The reason why $T_G$ 
does not appear to play a decisive role in the creation of topological defects 
is likely to be associated with the spatial extent of the 
thermally activated transitions:  Local thermal fluctuation can
perhaps create small defects, but these  topological 
defects will have a scale approximately
equal to the size of their core 
(since both are defined by the same  correlation length $\xi$). 
Such ill-defined defects are unlikely to survive --- after all, they  represent
a configurations which may be a shallow local 
minimum of the free energy, but which have a
higher  free energy than the uniform superfluid.  
Hence, they are not large enough to be really
topologically stable --- change of the field configuration 
in a finite region of space the order
of $\xi$ suffices to return to the uniform  ``true vacuum''.  
Moreover, such local defects cannot easily lead to the creation 
(or destruction) of the large ones.
These requirements are least convincing in the case of
monopoles (which have spatial  extent of order $\xi$ in every direction).  
However even
monopoles would have to be created ``in pairs'' by thermal  fluctuations.  
These pairs of
monopoles of opposite charge will be separated by distance of order $\xi$.  
Therefore,  they are not likely to separate and survive.

We are led to the conclusion that the dominant process in creation of topological
defects will have to  do with the critical slowing down and 
the subsequent ``freeze-out'' of the
fluctuations of the order parameter at the  
time $\hat{t}$ rather than with the thermal
activation and Ginzburg temperature.  

\section{Numerical Experiments}

Our numerical experiment \cite{Laguna97} 
to test the mechanism described in the previous section
for estimating the density of topological defects simulates the dynamics of
a real field $\Psi$ in a 1D system. This system contains all the essential ingredients
needed to investigate topological defect formation scenarios 
(including symmetry breaking,
although without an infinite range order),
and at the same time, because of its one-dimensional nature, it allows us to 
perform a reasonable and accurate numerical analysis.
We assume that the system under consideration is in contact with
a thermal reservoir and obeys the Langevin equation
\begin{equation}
\ddot \Psi + \eta \dot \Psi -
\partial_{xx} \Psi + \partial_{\Psi} V(\Psi) = \Theta \ .
\label{eq:motion}
\end{equation}
The potential $V$ is given by (\ref{eq:1}). (A rescaling of $x$ and $t$
can be introduced to eliminate the parameters $\lambda$ and $\sigma$.)
The noise term $\Theta$ has correlation properties
\begin{equation}
\langle \Theta (x, t), \Theta (x', t') \rangle =  2 \eta \theta \delta(x'-x)
\delta(t'-t) \ .
\label{eq:noise}
\end{equation}
In Eq.~(\ref{eq:noise}), $\eta$ is the overall damping constant 
characterizing the amplitude of the noise through Eq.~(\ref{eq:motion}). 
$\theta$ fixes the temperature of the reservoir.
Typical values used of $\theta$ were in the range ($0.1, 0.01$).

\begin{figure}[h]
\leavevmode
\epsfxsize=5.0truein\epsfbox{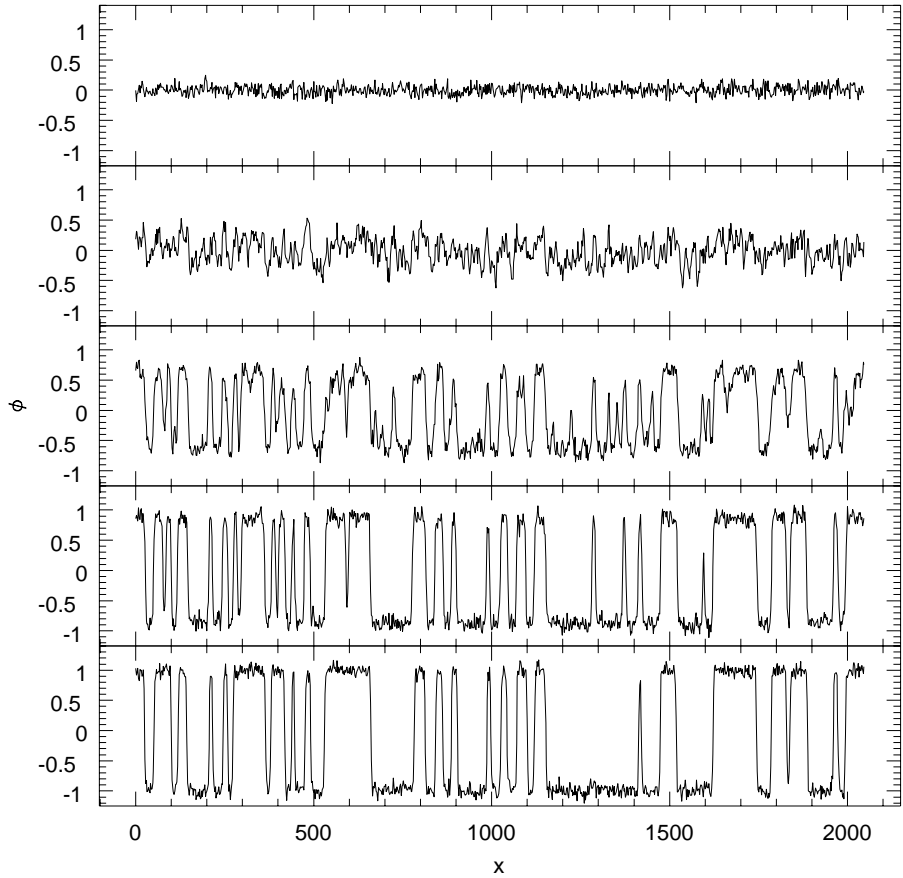}
\caption[figure2]{\label{figure2}
Snapshots of $\Psi$ during
kink formation with a quench timescale of $\tau_Q= 64$ and damping
parameter $\eta = 1$.
The figures, from top to bottom, correspond to $t$  = -80, 14, 33,
51 and 301, respectively (compare with \cite{Antunes97}).}
\end{figure}

Since the symmetry in the system is discrete ($\Psi \rightarrow -\Psi$), 
our numerical experiment consisted of investigating the creation of kinks 
(domain walls in 1D) as a function of the quench timescale $\tau_Q$.
Figure\ \ref{figure2} shows a sequence of
``snapshots'' of $\Psi$ for a typical quench. 
One can clearly see from this figure the transition from expectation
value $\langle\Psi\rangle = 0$ above the critical temperature ($\epsilon < 0$) to
$ \langle\Psi\rangle = \pm \sqrt { \epsilon } $ 
after the symmetry has been broken ($\epsilon > 0$).

\begin{figure}[h]
\leavevmode
\epsfxsize=5.0truein\epsfbox{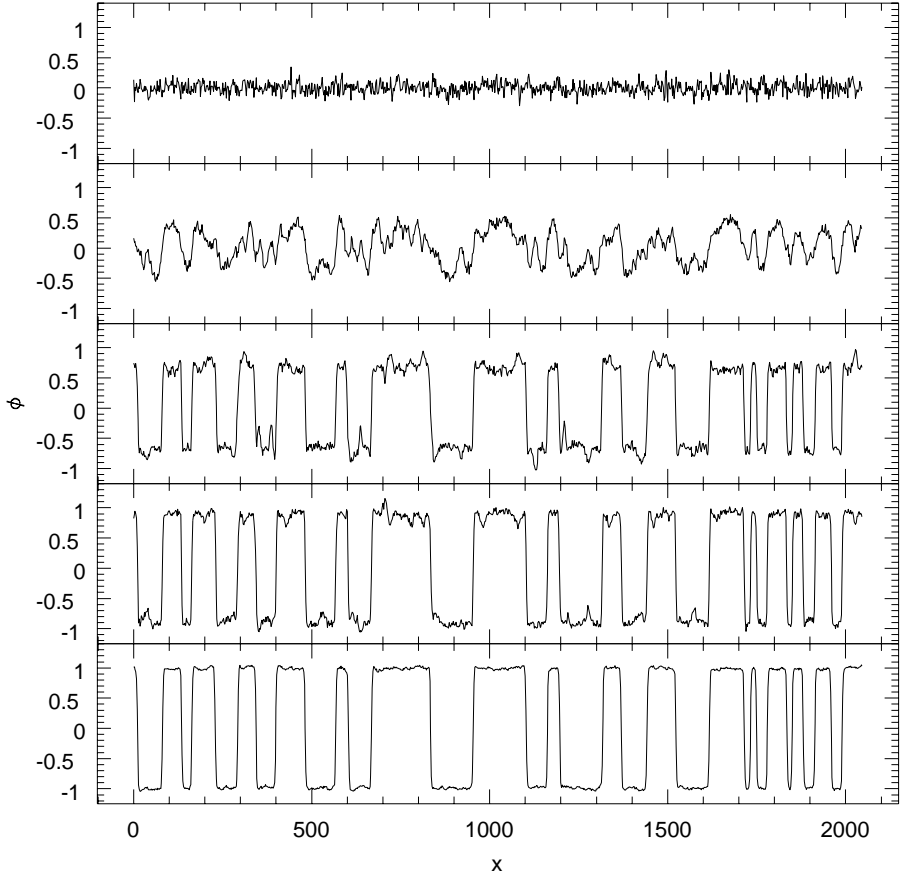}
\caption[figure2a]{\label{figure2a}
Snapshots of $\Psi$ during
kink formation with a quench timescale of $\tau_Q= 64$ and damping
parameter $\eta = 1/64$.
The figures, from top to bottom, correspond to $t$  = -80, 14, 33,
51 and 301, respectively. As expected from Eq.~(\ref{eq:xi2}),
the number of kinks is approximately half of what is seen in
Fig.~\ref{figure2}, where $\eta = 1$}
\end{figure}

\begin{figure}[h]
\leavevmode
\epsfxsize=5.0truein\epsfbox{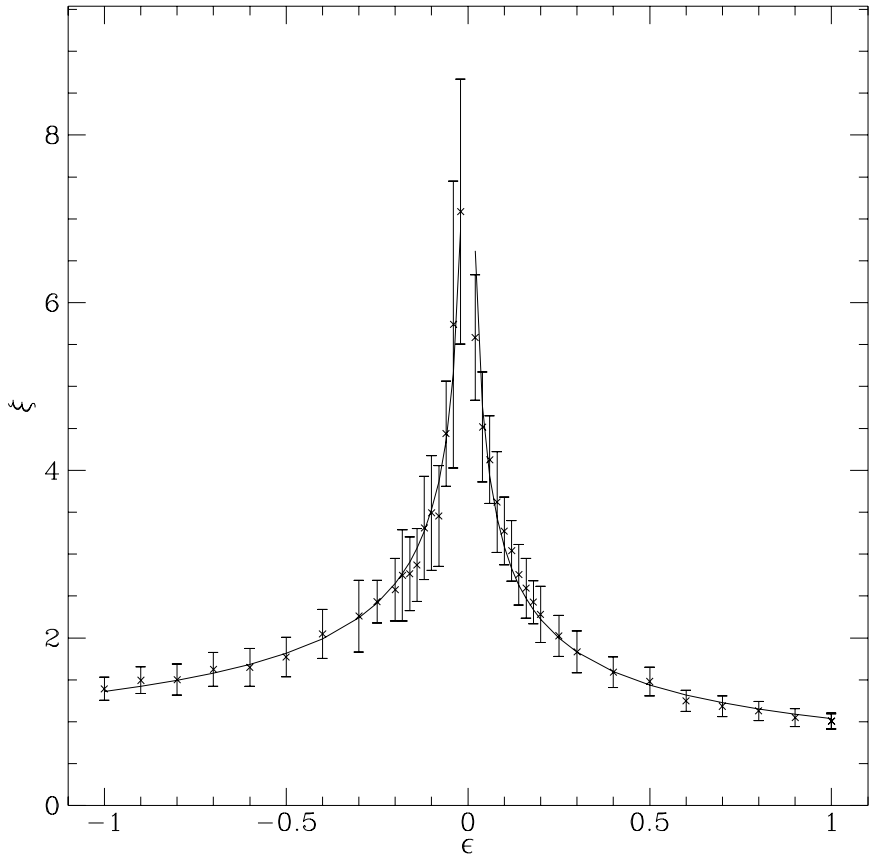}
\caption[figure3]{\label{figure3}
Equilibrium correlation length with solid lines
representing a fit to
$\xi = \xi_0 | \epsilon |^{-\nu}$.
The best fitting 
yields $\xi_0 = 1.38 \pm 0.06$, $\nu = 0.41 \pm 0.03$ ($\chi^2 = 1.2$) above $T_C$,
and  $\xi_0 = 1.02 \pm 0.04$, $\nu = 0.48 \pm 0.02$ ($\chi^2 = 3.7$) below $T_C$,
close to the Landau-Ginzburg exponent of $\nu = 1/2$.}
\end{figure}

We test the predictions of defect density in \cite{Zurek85b} 
in the overdamped regime ($\eta = 1$),
that is, the regime for which 
the damping term $\eta \dot \Psi$ in Eq.~(\ref{eq:motion}) is bound to dominate
over $\ddot \Psi$.
For comparison, we also show in Figure~\ref{figure2a} a sequence of
``snapshots'' of $\Psi$ for an underdamped case with $\eta = 1/64$.
Overdamped evolution will take place whenever
$ \eta^2 > | \epsilon |$ (see e.g. \cite{Antunes97}).
The characteristic relaxation time is then given by
$ \tau = \eta / |\epsilon| \ , $ and therefore, from Eq.~(\ref{eq:tau}),
we have that $\mu = 1$ and $\tau_o \approx \eta$.
Furthermore, from Eq.~(\ref{eq:ehat}), one obtains that the freeze-out
relaxation time in this case is
\begin{equation}
\hat t = \sqrt{ \eta \, \tau_Q} \ .
\label{eq:cond1}
\end{equation}
One can then use (\ref{eq:cond1}) to estimate from (\ref{eq:ehat}) and (\ref{eq:xihat})
both, the freeze-out temperature parameter $\hat \epsilon$ and correlation length
$\hat \xi$. 
Their values are
\begin{equation}
\hat \epsilon = \hat t / \tau_Q = \sqrt{\eta / \tau_Q} 
\end{equation}
and
\begin{equation}
\hat \xi = \xi_0 ~(\tau_Q/\eta)^{1/4} \ ,
\label{eq:xi2}
\end{equation}
respectively, where we have adopted $\nu = 1/2$ 
in accord with the Landau-Ginzburg theory.
The value of $\nu = 1/2$ is also in agreement with
the calculated critical exponents of the correlation length 
inferred from our equilibrium simulations (see Fig.\ \ref{figure3}).
We obtained for the equilibrium correlation length a best fit yielding
$\xi_0 = 1.38 \pm 0.06$, $\nu = 0.41 \pm 0.03$ above $T_C$,
and  $\xi_0 = 1.02 \pm 0.04$, $\nu = 0.48 \pm 0.02$ below $T_C$,
close to the Landau-Ginzburg exponent of $\nu = 1/2$.

\begin{figure}[h]
\leavevmode
\epsfxsize=5.0truein\epsfbox{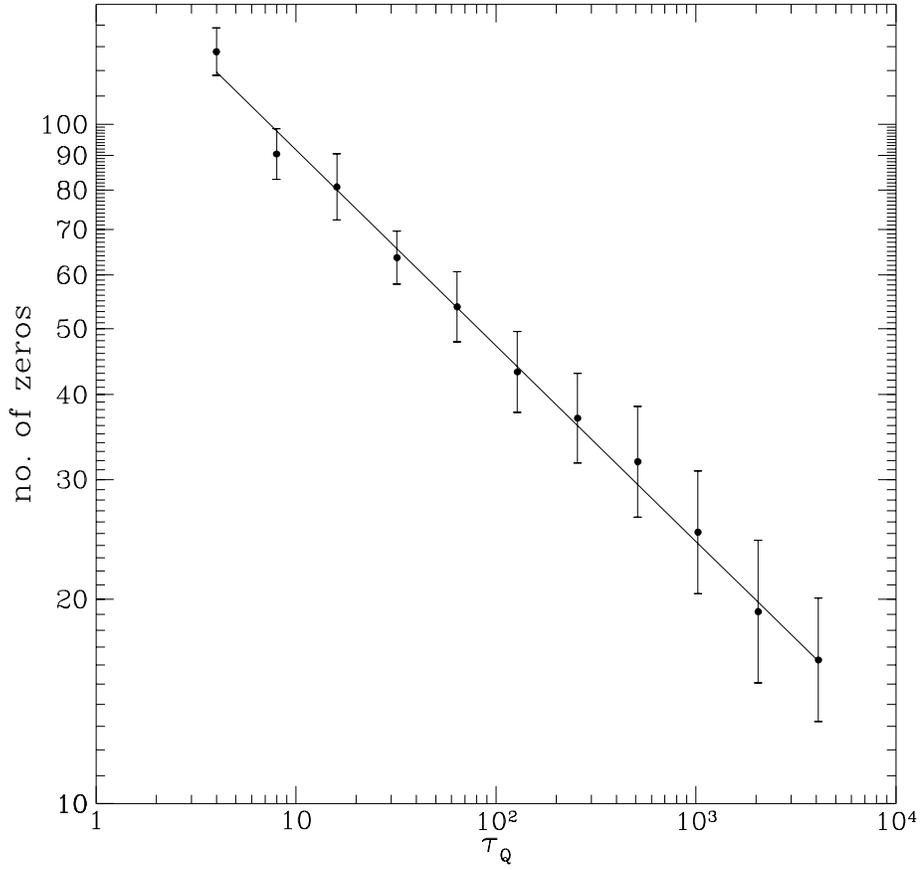}
\caption[figure4]{\label{figure4}
Number of defects as a function of quench timescale.
The number of kinks is calculated at a time $t/\tau_Q = 4$.
The straight line is the best fit to $N = N_o\,\tau_Q^{-a}$
with $a = 0.28 \pm 0.02$ and $N_o = 178 \pm 14$ ($\chi^2 = 1.96$).}
\end{figure}

\begin{figure}[h]
\leavevmode
\epsfxsize=5.0truein\epsfbox{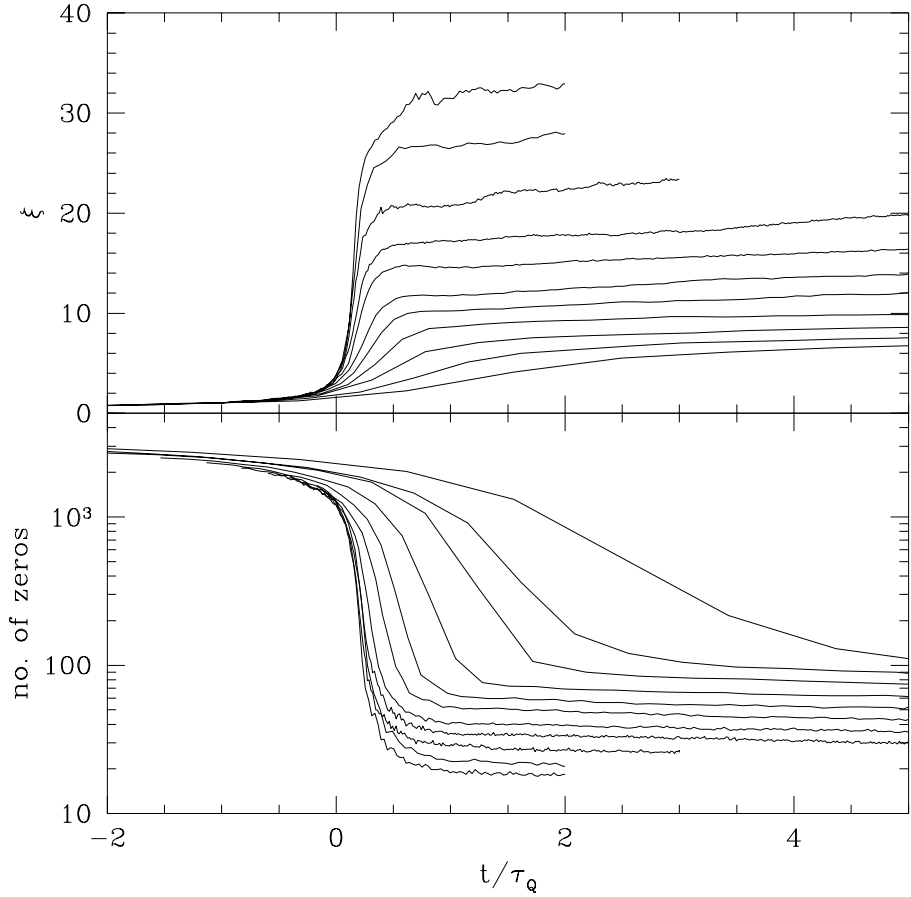}
\caption[figure5]{\label{figure5}
Correlation length $\xi$ and average number of zeros 
as a function of time in units of $\tau_Q$;
from bottom to top for $\xi$ and
top to bottom for the average number of zeros, 
$\tau_Q = 4, 8, ..., 2048, 4096$.
The number of kinks used in Fig.~\ref{figure3} were
obtained at $t/\tau_Q \sim 4$, except in the large, computationally expensive,
$\tau_Q$ cases where an extrapolated value was used.}
\end{figure}

Figure\ \ref{figure4} encapsulates the main result of our paper. 
It shows the number of zeros (kinks)
obtained in a sequence of quenches with various values of the quench
timescale $\tau_Q$.
For each value of $\tau_Q$, 15 quenches were performed in 
a computational domain of size 2,048 units with periodic boundary conditions.
This size of computational domain is large enough to avoid boundary effects.
The number of kinks produced after the quench was obtained by counting
the number of zeros of $\Psi$. 
Above and immediately below $T_C$,
there is a significant number of zeros which have
little to do with the kinks (see Fig.\ \ref{figure2}).
However, as the quench proceeds, the number of zeros quickly evolves towards
an ``asymptotic'' value
(see Fig.\ \ref{figure5}). This change of the density of zeros and
its eventual ``stabilization" is associated with
the obvious change of the character of $\Psi$ and with
the appearance of the clearly defined kinks.
By then, the number of kinks is nearly constant in the runs with long
$\tau_Q$, and, correspondingly, their kink density is low. Even in the runs
with the smallest $\tau_Q$, there is still a clear break
between the post-quench rates of the disappearance of zeros and the long-time,
relatively small rate at which the kinks annihilate.
(Kinks will eventually annihilate, but on slow timescales
set by the thermal diffusion of kinks over distances of the order 
of their separation.)

Our simulation clearly indicates \cite{Laguna97} a
scaling relation for the number density of kinks.
We find a kink density (no. of zeros/2,048) of
$n = (0.087 \pm 0.007)\, \tau_Q^{-0.28 \pm 0.02}$,
when the kinks are counted at approximately the same $t/\tau_Q$ value.
By contrast, when the counting is done at the same value of
absolute time $t$, $n$ has a somewhat shallower 
slope ($n \propto \tau_Q^{-0.24}$), reflecting
higher annihilation rate at high kink densities.
Within the estimated errors, this result is in agreement with the
theoretical prediction of an scaling of 
$ n \approx (\eta/\tau_Q)^{1 \over 4}$.

An important point to consider is that there 
are of course no ``true" phase transitions in
1D, and that instead of vortices or strings we are counting kinks. 
The remarkable agreement of our numerical results \cite{Laguna97} with the
theoretical predictions suggests that the crucial ingredient 
in the scenario of Ref.~\cite{Zurek85b} 
is not that much the existence of a ``mathematically rigorous" phase transition.
The essential conditions required for this scenario are:
the presence of symmetry breaking and the
scalings (\ref{eq:xi}) and (\ref{eq:tau}) for the
correlation length and relaxation time.

\section{Summary and Discussion}

The central issue of the work presented in this paper was a numerical test
of the mechanism for defect formation in the course of second order phase
transitions involving a non-conserved order parameter. 
Both laboratory \cite{Hendry94,Ruutu96,Bauerle96}
and computer experiments \cite{Laguna97}
have added strong support
to the proposal that the initial density of the topological defects will be set by the 
correlation length at the freeze-out instant $-\hat t$; that is, at the moment
when the relaxation timescale of the order parameter will be comparable to the
characteristic timescale of the quench.
At that instant, as a consequence of critical slowing down, perturbations of 
the order parameter become so sluggish that they cease to keep up with the
equilibrium scalings, 
so that in the time interval [$-\hat t, \hat t$] the order parameter cannot
adjust to the changes of thermodynamic parameters induced by the quench.
By the time $t > \hat t$ below $T_C$, 
the dynamics will ``restart''; however, it will be too late for the system to get 
rid of the defects, by then the defects will be ``set in concrete,''
that is, stabilized by the topology (Although subsequent evolution 
involving their interactions will ensue, and, generally alter their density). 

Strong evidence that thermal activation is not responsible for the
formation of topological defects is provided by the 
experimental results obtained in the superfluid He$^4$ experiment by the Lancaster
group \cite{Hendry94} and by the He$^3$ experiments in
Helsinki and Grenoble \cite{Ruutu96,Bauerle96}.
Together, they demonstrate that the argument 
in Ref.~\cite{Zurek85b} is unaffected by 
the thermal activation process invoked in Kibble's original discussion 
(\cite{Kibble76}, see also \cite{Gill96}).
Ginzburg temperature --- that is, the temperature at which vacuum can 
be thermally ``flipped'' over the potential barrier in regions of size $\xi$ ---
is far below the critical temperature in the superfluid helium.

Our numerical experiments 
also provide a strong confirmation of the
theoretical predictions of the non-equilibrium scenario \cite{Zurek85b}.
The scaling of the kink density with $\tau_Q$ follows closely
theoretical expectations. These results are also supported by the dependence
of the number of kinks on the damping parameter $\eta$
(compare Fig.~\ref{figure2} and Fig.~\ref{figure2a}), as well as by the
preliminary results of the computer experiments involving complex order
parameter and/or more than 1D space that we are currently investigating.

\subsection*{Acknowledgments}
We thank Yu. Bunkov, S. Habib, T.W.B. Kibble and M. Krusius for helpful discussions.
Work supported in part by 
NSF PHY 96-01413, 93-57219 to P.L. and
NASA HPCC to W.H.Z.

%
%        figure 1
%

%
%        figure 2
%

%
%        figure 2a
%

%
%        figure 3
%

%
%        figure 4
%

%
%        figure 5
%

\end{document}